\documentclass[twocolumn,showpacs,preprintnumbers,amsmath,amssymb]{revtex4}

\usepackage{graphicx}
\usepackage{dcolumn}
\usepackage{bm}

\begin{document}
\title{How colloidal suspensions that age are rejuvenated by strain application?}

\author{Virgile Viasnoff}
\affiliation{
L.P.M UMR7615 CNRS ESPCI 10 rue Vauquelin 75231 Paris, France
}
\author{St\'ephane Jurine}
\affiliation{
L.P.M UMR7615 CNRS ESPCI 10 rue Vauquelin 75231 Paris, France
}
\author{Fran\c{c}ois Lequeux}
\affiliation{
L.P.M UMR7615 CNRS ESPCI 10 rue Vauquelin 75231 Paris, France
}

\date{\today}
\begin{abstract}
We present here a microscopic study of the effect of shear on a
dense purely repulsive colloidal suspension. We use Multispeckle
Diffusing Wave Spectroscopy to monitor the transient motions of
colloidal particles after being submitted to an oscillatory
strain. This technique proves efficient to record the time
evolution of the relaxation times distribution. After a high
oscillatory shear, we show that this distribution displays a full
aging behavior. Oppositely, when a moderate shear is applied the
distribution is modified in a non trivial way. Whereas high shear
is able to erase all the sample history and rejuvenate it, a
moderate shear helps it to age. We call this phenomena
\textit{overaging}. We demonstrate that overaging can be
understood if the complete shape of the relaxation time
distribution is taken into account. We finally report how the Soft
Glassy Rheology model accounts for this effect.

\end{abstract}

\pacs{07.60.-j, 78.35.+c, 81.40.Cd}

\maketitle

\section{Introduction}
The study of dense colloidal suspensions still raises many open
questions. One interesting feature is that some of theses systems
remain out of equilibrium for any accessible experimental time.
For instance, if the pair potential of interaction has an
attractive part, the particles can aggregate in an always evolving
physical network \cite{cipelletti00}. If the potential is purely
repulsive, an increase of concentration leads to a jamming
transition and the suspension can reach an amorphous state or
glassy phase \cite{PuseyVanMeganColl}. More precisely, in this
last case the characteristic relaxation time of the system
increases dramatically around
  a critical value of
   the volume fraction $\phi_c$.  For large enough concentrations, the relaxation processes rapidly become longer
   than any experimental time. The system is then macroscopically pasty and can display a non-stationary behavior.
    A direct consequence of the extremely long relaxation
    times is that all the past history of the sample has to be taken into account. A key trick to obtain reproducible
    results is to perform experiments on a sample with the exact same history. As far as rheological measurements are
     concerned, it is known for long that a high preshear can erase all past memories in many systems. It is thus
     often used as a trick to reset the sample history provided that it does not damage it irreversibly. This
     procedure is reminiscent of the process of thermal quench used to erase the history for other structural
     or spin glasses \cite{Revueverrespin,Struikbook}. This similarity let envision a comparable role played
     by shear and temperature on a microscopic level. It has recently been shown on different 'pasty' colloidal
     suspensions that their macroscopic behavior after a rheological quench  have the same qualitative feature
     than other glasses after a temperature quench \cite{CaroFranc,cloitre1}. However, measurements of the effect of shear
     on the dynamic at a microscopic level are still lacking. In this paper we study a colloidal glass with Multispeckle
     Diffusing Wave Spectroscopy (MSDWS), monitoring the motions of the colloidal particles after that the sample
     underwent various strain histories. We will show that the mechanical perturbation acts in a dual fashion on
     the microscopic dynamic. All the observed effects will be discussed within the Soft Glassy Rheology (SGR)
     model \cite{PRLLeqHeb,solirevue,Sollitchseul}.

\section{Background}
In this section we will recall some general results on glasses.  We will mainly focus on results obtained on spin and
 structural glasses upon a change of temperature. However, most of theses results are similar to that observed in
  colloidal glasses if a temperature decrease is substituted by an increase of
  concentration \cite{PuseyVanMeganColl,Hartl}.
Spin and molecular glasses display a qualitative change in their
microscopic dynamic when the temperature is lowered around the
glass transition temperature.
  Actually, the distribution of relaxation times splits into two distinct families
  when the temperature is decreased (see eg :\cite{PuseyVanMeganColl,Larson}): on the one hand, short
   distance motions and vibrations of the particles can be describe by fast
   individual modes called $\beta$ modes. Their sensitivity to temperature
   changes
   is weak. On the other hand collective motions i.e. structural relaxations,
   occur through a broad distribution of very slow modes called $\alpha$ modes.
   This modes are very sensitive to temperature changes. In general, the characteristic
    time of the $\alpha$ relaxation exhibits such a huge increase when the temperature
     is lowered, that it seems to diverge, up to experimental evidence.
        For low enough temperatures, such that the relaxation time is larger than the experiment time by
        many orders of magnitude, the systems present
      some strange non-equilibrium features. One
      of the most striking among them, is the so called aging phenomena. It consists
      in a drift of the $\alpha$ relaxation time distribution, towards longer and
      longer relaxation times. This distribution keeps on evolving for any experimental
       time scale. The simplest way to characterize this drift is to perform a quench
       from a high temperature state where an equilibrium distribution can be achieved
       to a low temperature state in the glassy region \cite{Struikbook,Revueverrespin}.
       In this case, the age of the system is defined as the time elapsed since the quench,
       i.e. the time spent in the glassy phase. In theses conditions, it has been proven
       theoretically and experimentally alike that any measured quantities depends explicitly on the age of the system
         \cite{Revueverrespin}. More precisely, the response and correlation functions
         do not only depend on the elapsed time $t$ since the beginning of the measurement as
          it would be the case for an equilibrium system but they also depend on the age of
           the system $t_w$, at which the measurement started. Consequently, any correlation
            function $g$ writes $g(t_w,t+t_w)$. Predictions and experimental results show that,
             despite the absence of equilibrium, there is a regime  often called asymptotic regime,
             where the only relevant time scale is the very age of the system. As a consequence
             response or correlation functions can be rescaled with $t_w$.
             One finds that  $g(t_w,t+t_w)=g(\frac{h(t+t_w)}{h(t_w)})$ where h is a function
             that depends on the system \cite{BerthierBouchaud}. When $g(t_w,t+t_w)=g(\frac{t}{t_w})$,
              the scaling is usually called full aging. We point out, that the existence of such a rescaling
              in this asymptotic regime shows that the distribution of relaxation times drifts in a self-similar way. Such an asymptotic regime can only be reached if the system is left at a constant temperature for a sufficiently long time in the glassy phase. However, if some additional energy is transiently provided to the aging system, the dependance of the
relaxation functions with the age of the system becomes extremely
intricate. For example, temperature ramps with stops lead to the
very remarkable "memory effect" as observed in spin glasses - see
eg \cite{memoeff} - and recently in polymer glasses
\cite{Cilibertopolymem}. The shape of the relaxation times distribution
is then modified in a non-trivial way during potentially very long
transient regimes. The analysis of this behavior would help to
understand the precise modification of the relaxation time distribution  by
external parameters, and thus to get an insight of the system's internal dynamics .
This is the scope of this paper. We have chosen to work
on a colloidal glass where it is easy to measure a time-dependant
correlation function.
\newline For colloids, temperature is not a practical parameter. But since it
is believed that temperature and shear may act similarly in theses
systems \cite{LiuNagel}, we studied the influence of the shear
upon the microscopic dynamics of a colloidal suspension.
\newline In this paper we will focus on the modification of the shape of the distribution
of $\alpha$ relaxation times $P(\tau_\alpha)$. We will show that
injecting transiently some mechanical energy into the system
modifies the distribution of relaxation times in a dual fashion.
Conversely to what could be intuitively expected, it can both
rejuvenate or \textit{overage} the system. In practice, we study a
dense suspension of purely repulsive polystyrene beads both
electrostatically  and sterically  stabilized. We record the
fluctuations of a laser light, multiply scattered by the sample.
We compute the two times intensity autocorrelation function
$g_2(t+t_w,t_w)$. The variation of $g_2$ is directly related to
the mean square displacement of the scatterers. In order to
"quench" the system properly, the sample is first presheared by a
large oscillatory strain. When the shear is stopped we record
$g_2(t+t_w,t_w)$ for evenly spaced $t_w$. We show that it displays
the properties of full aging. Then the sample is   submitted to a
burst oscillatory strain of different amplitude and frequency. We
monitor the change in the shape of $g_2$ and interpret it in terms
of changes in the distribution of relaxation times.

\section{Sample preparation and Experimental setup}
The sample is a commercial suspension of polystyrene spherical
beads of diameter 162 nm copolymerized with acrylic acid (1\%)
that creates a charged corona stabilizing the microspheres. The
corona prevents both from aggregation by steric and electrostatic
repulsions and from crystallization. The suspension was carefully
dialyzed to a polymer volume fraction of $\phi=50\%\pm 0.5\%$. The volume
fraction was determined by drying and was chosen to fulfill the
two following criteria:
\begin{description}
  \item[-] The sample has to be concentrated enough so that both $\alpha$ and $\beta$ modes exist.
  The amplitude of the $\beta $ mode must be sufficiently small so that the alpha mode can be accurately measured.
  The $\alpha$ modes must also display some aging behavior over all the experiment time scale.
  \item[-] The volume fraction has to be low enough so that the strain is homogeneous, at least on macroscopic length scale all through the sample. We checked that no effect like shear banding occurs for the concentrations in use.
\end{description}
Because of the divergence of the viscosity with the volume fraction
in the vicinity of the glass transition, the suitable range of concentration fulfilling these two requirements is narrow (between 50\% and 52\%).

 \begin{figure}[ht]
\begin{center}
\includegraphics[width=8cm,keepaspectratio=true]{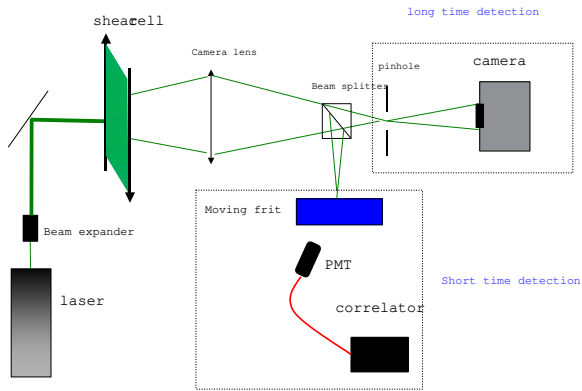}
\caption{Experimental set-up. The sample is placed in a shear
cell.The emerging light is simultaneously analyzed by the fast
correlator and the camera. The problem of ergodicity is solved by
performing a spatial average for the camera part, and moving a
frit glass in front of the PMT detector for the fast detection.}
\label{montage}
\end{center}
\end{figure}

Fig \ref{montage} shows the experimental setup. The sample is illuminated using polarized light from an argon-ion laser
operating at a wavelength of $\lambda = 514$~nm. The laser beam is expanded
to a diameter of approximately 1~cm and is incident on the sample cell. Multiply scattered light is
collected by a 50-mm Nikon camera lens and split into two parts. The first part is focused onto an iris diaphragm.
The lens is set up so that a one-to-one image of the scattered light from
the output plane of the sample cell is formed on the diaphragm. The scattered light emerging from the diaphragm is detected by a Dalsa CCD camera, model CAD1-128A,
which is placed approximately $d\simeq 15$~cm behind the diaphragm. It is an
8-bit camera which we run at 500 frames per second. The images are
transferred to a computer running at 500~MHz using a National Instruments
data acquisition board, model PCI1422. The combined speeds of the computer
and data acquisition boards are sufficiently high to allow data to be
analyzed in real time. The accessible times range from $10^{-3}$s to $10^4$ s.
\newline The second part of the light is shone onto a moving frit. Its motion is controlled by a piezo actuator. The light emerging from the frit is detected by an optical fiber, amplified by an ALV photomultiplier tube (PMT) and analyzed by a Flex correlator. The moving frit has the same effect as the second cell in the two cells technique \cite{romer00} but we found it easier to implement. It allows an exact determination of the non ergodic correlation function at very short times $(10^{-8}s,10^{-1}s)$. A complete description of the technique can be found in \cite{Experimental}. In the transmitted geometry the correlation functions calculated with the camera and with the PMT can be overlaid \cite{Experimental,MicelleScheff}. The dynamics of the beads can thus be probed over 12 decades in times.
The sample is placed in a custom-made shear
cell consisting in two parallel glass plates with a variable gap. For all
presented experiments, the gap was set to 1.3 mm. Oscillatory straining was
realized by moving the bottom plate thanks to a piezoelectric device. Shear
strain from 30\% to 0.04\% could be possibly applied at different frequencies ranging from 0.01Hz to 10Hz.

\section{Experiments}

\subsection{rheological quench}
\begin{figure}[ht]
\begin{center}
\includegraphics[width=8cm,keepaspectratio=true]{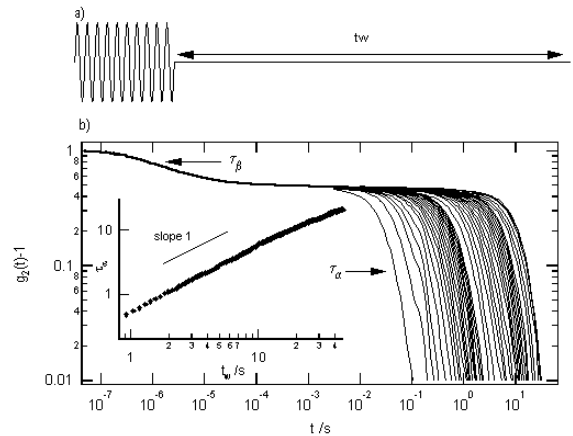}
\caption{Normalized intensity autocorrelation functions. The fast
decrease of the function corresponds to the $\beta$ modes and is
measured via the correlator. The long time decay corresponds to
the $\alpha$ modes -structural rearrangements. It never reaches
any steady state on our experiment time scale. The $\beta$ modes
are insensitive to shear and are in a steady state. For sake of
clarity only one curve was plotted for the short times. Inset:
$\tau_\alpha$ versus $t_w$. We find that the relaxation time
varies linearly with the waiting time.} \label{superposition}
\end{center}
\end{figure}

In order to check if an oscillatory strain is able to entirely rejuvenate the sample, we submitted the suspension to a series of shear strain of different amplitudes at a fixed frequency of 1Hz for 100s. The measurement of the correlation function starts when the strain is stopped (see fig \ref{superposition}a). The shear cessation is taken as the origin for the age $t_w$ of the system. For a strain amplitude above 20\% the correlation functions become insensitive to the strain amplitude and to the past history. A reproducible state is reached. We have checked that this state does not depend on the duration of the strain application provided that it is superior to 40s. Typical curves are plotted on fig \ref{superposition}b for a volume fraction of 51\%.

\begin{figure}[ht]
\begin{center}
\includegraphics[width=8cm,keepaspectratio=true]{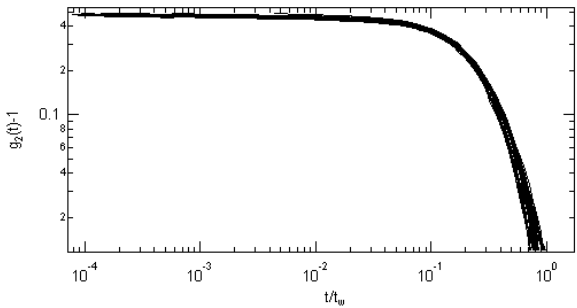}
\caption{$\alpha$ relaxation of the intensity autocorrelation
function vs $\frac{t}{t_w}$. The rescaling of the curves is
satisfactory.} \label{rescale}
\end{center}
\end{figure}

The correlation function shows a characteristic two steps decay as earlier mentioned. A first decrease happens around $\tau_\beta=10^{-6}$s. It is insensitive to shear (for sake of clarity only one curve has been plotted). Then the correlation function plateause at a value $g_{plat}\simeq 0.5$. We use a regular algorithm to extract the mean square displacement of the beads for this plateau value. It corresponds fluctuation of position over a distance of  $\delta \simeq 9nm$. Finally a second decrease takes place at a time $\tau_\alpha$ that varies with the age of the system. $\tau_\alpha$ is arbitrarily defined so that $g(t_w, t_w+\tau_\alpha)= \frac{1}{2}g_{plat}$. The inset of fig \ref{superposition} shows that $\tau_\alpha$ evolves proportionally with the sample's age $t_w$. All the aging part of the curve can be rescaled by $t/t_w$ as displayed on fig \ref{rescale}. The aging part displays thus the characteristic scaling of full aging. It means that $P(\tau_\alpha)$ evolves in a self-similar manner. The high shear provided the system enough energy to entirely rejuvenate it. It thus corroborates the macroscopic observations previously mentioned. It also emphasize the analogy between temperature and strain as far as quenches are concerned.

\subsection{Effect of a moderate oscillatory shear}
\begin{figure}[ht]
\begin{center}
\includegraphics[width=8cm,keepaspectratio=true]{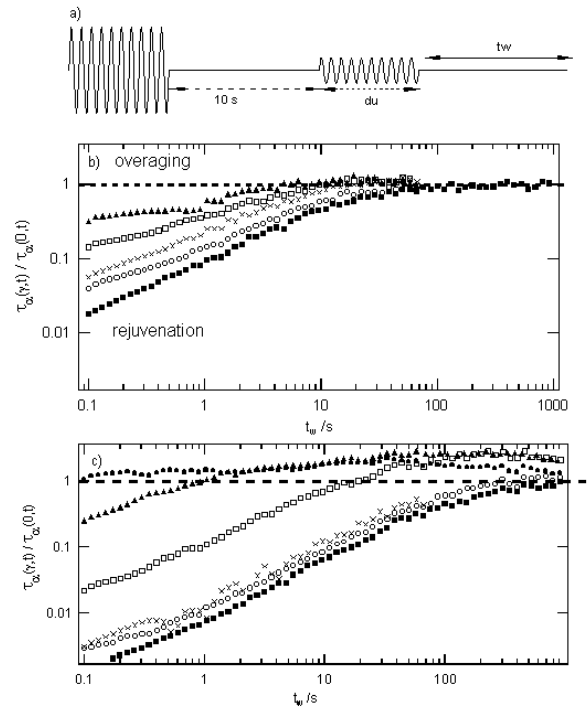}
\caption{a) Strain history $\gamma(t)$. We vary both the amplitude
and duration of the second burst. b) Normalized relaxation time
$\tau_\alpha(\gamma_0,t_w)/\tau_\alpha(0,t_w)$ vs $t_w$ after a
burst of duration 1s at 1Hz for different amplitudes
$\gamma_0=2.9\%(\bullet$),$\gamma_0=5.9\%$({\tiny
$\blacktriangle$}),$\gamma_0=7.9\%$({\small
$\square$}),$\gamma_0=11.7\%$($\times$),$\gamma_0=14.5\%(\circ)$,
and complete rejuvenation ({\tiny $\blacksquare$}). c) Same curves
for a duration of 100s.Notice that for the lowest shear amplitude
overaging occurs. } \label{taualpha}
\end{center}
\end{figure}

We now examine the influence of the strain amplitude upon the
microscopic dynamics. All the following curves are taken in
backscattering geometry for a volume fraction $\phi=50\%$ and a
measured value of $\delta \simeq 68nm$. As demonstrated in the
previous section, a high shear is able to entirely rejuvenate our
system. Hence we first submit the sample to an oscillatory strain
of 30\% for 100s in order to obtain reproducible results.
Secondly, the system is left at rest for 10s then it is submitted
to a second burst of oscillatory strain of different amplitude
$\gamma_0$ and duration $du$. We set the frequency to 1Hz. The
strain history is displayed on fig \ref{taualpha}a. We now take
the origin of the system's age just after the \textbf{second}
burst. This convention is purely arbitrary but it allows an easier
representation of our results on a log scale. We record
$\tau_\alpha(\gamma_0,t_w)$ as a function of $t_w$ for different
amplitudes of $\gamma_0$. We take the curve for $\gamma_0=0\%$ as the
reference curve. On fig 3b we plotted the ratio
$R=\tau_\alpha(\gamma_0,t_w)/\tau_\alpha(0,t_w)$. If $R<1$, then
the effect of shear is to rejuvenate the system, that is to say
that the sheared sample has a quicker dynamics than the
unperturbed one. If $R>1$, then the internal dynamics is slowed
down by the shear application. We call this situation "overaging".
If a single oscillation is applied at 1Hz for 1s, the sample is
partially rejuvenated whatever the strain amplitude may be. This
is shown on fig \ref{taualpha}b where all the curves lie in the
region where $R<1$. One can wonder if the partial rejuvenation
corresponds to a simple backward shift of $P(\tau_\alpha)$. In
other words, does the rejuvenation process under shear is the time
reversal process of aging? If this would be the case, one could
define an effective age $t_{eff}$ such that
$\tau_\alpha(\gamma_0,t_w+t_{eff})=\tau_\alpha(0,t_w)$.

 Experimentally we could not find such an effective age and time translation proved inefficient to collapse all the $\tau_\alpha(\gamma_0,t_w)$ onto a master curve. It thus mean that $P(\tau)$ is not modified by the strain in a self-similar manner. However, at long $t_w$ all the curves merges to $R=1$, indicating that at long times the self-similarity is recovered as expected.
\par However if 100 oscillations are applied for 100s at 1Hz, the situation is changed as shown on fig \ref{taualpha}c. For the highest shear amplitude the ratio R remains constantly below 1 consistently with the results of the previous section. However for the smallest strain amplitudes, R is first inferior to one but then becomes superior as $t_w$ increases. The effect of shear for this amplitude is thus dual. Shortly after the burst the dynamics is accelerated. But after a while it becomes slower than that of the unperturbed case. It thus means that the transient shear application has modified the distribution of the relaxation times not only in a non self-similar manner but also in a non monotonic way.
\newline Notice, however that at very long times all the curves seem to converge to R=1. It shows that regular aging is recovered.
By considering only $\tau_\alpha$ in the relaxation process we have reduced the relaxation to a single time. However, the shape of the correlation function is the result of the whole distribution of relaxation times. It is thus interesting to compare the full correlation function $g_2(t,t_w,\gamma_0)$ to the reference one $g_2(t,t_w,0)$.

\begin{figure}[ht]
\begin{center}
\includegraphics[width=8cm,keepaspectratio=true]{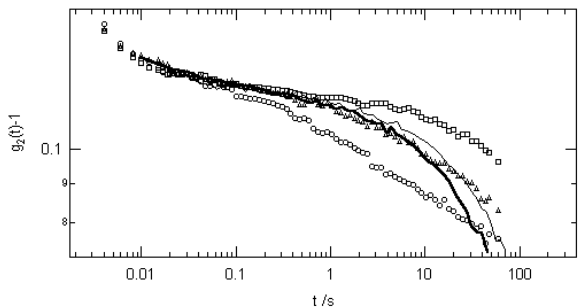}
\caption{Correlation functions obtained after the second burst in
the reference case  for $t_w=10^{-1}s$ (Bold line) $t_w=60s$ (Grey
Line) and in the case $\gamma_0=5.9\%$ for $t_w=10^{-1}s$ (O),
$t_w=1s$ ($\triangle$), $t_w=60s$ ($\square$). Notice the crossing
of the curves occurring around 30s. The change in the shape of the
correlation function is clearly visible.} \label{functcorr}
\end{center}
\end{figure}

 Fig \ref{functcorr} shows the reference curves $g_2(t,t_w,0)$ for $t_w=0.1s$ (dark line) and $t_w=60s$ (grey line). We emphasize the fact that $t_w$ is now referenced from the cessation of the burst. The symbol curves represent $g_2(t,t_w,7.9\%)$ for $t_w=0.1s$ ($\circ$), $t_w=1s$ ({\tiny $\triangle$}), and $t_w=60s$ ({\tiny $\square$}). The comparison of the curves for $t_w=0.1s$ reveals that $g_2(t,t_w,7.9\%)$ starts decreasing earlier than $g_2(t,t_w,0\%)$. It is thus an indication that faster relaxation process occurs in the sample. However, at long times, $g_2(t,t_w,7.9\%)$ crosses and then lies above $g_2(t,t_w,0\%)$. It thus mean that at long times the relaxation processes in the sheared sample are slower than in the reference one. This is confirmed by the relative position of $g_2(t,t_w,7.9\%)$ and $g_2(t,t_w,0\%)$ for $t_w=60s$. $g_2(t,t_w,7.9\%)$ decreases more slowly than $g_2(t,t_w,0\%)$. Thus, the fast relaxation times have aged and the dynamics is dominated by the slower ones. The change in the shape of the correlation function is the very sign that the relaxation time distribution has not been simply shifted backwards in time by a constant amount, but that its shape has been modified with an addition of short and long relaxation times.

\subsection{effect of frequency}
\begin{figure}[ht]
\begin{center}
\includegraphics[width=8cm,keepaspectratio=true]{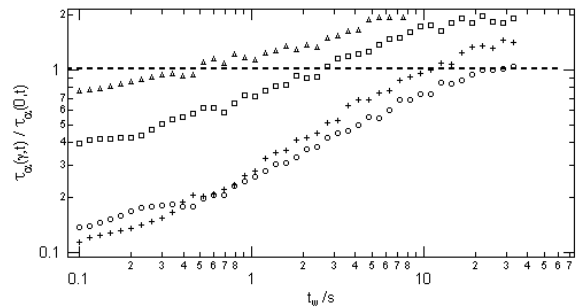}
\caption{Normalized $\alpha$ relaxation time for $\gamma_0=7.9\%$
at various frequencies: 0.1Hz($\circ$), 1Hz(+), 5Hz ({\tiny
$\square$}), 10Hz ({\tiny $\triangle$}). Notice that the
importance of the overaging regime increases with the frequency. }
\label{freqexp}
\end{center}
\end{figure}

Up to now, we have only focused on the amplitude and duration of the second burst. However, one can also expect that its frequency plays a role in the rejuvenation-overaging process. Indeed, one could expect that not only strain but also strain rate is an important factor in this mechanism. We set the amplitude to 7.9\% and the duration of the burst to 10s. We varied the frequency from 0.1Hz to 10Hz. Hence the sample was submitted to a number of oscillationsn ranging from 1 to 100. The result are displayed on fig \ref{freqexp}. Notice that the overaging effect is all the more pronounced that the frequency is high. We believe though that the overaging behavior cannot be observed under steady strain rate as usually performed in rheological measurements. However, we could not check this hypothesis since our set up does not allow continuous straining.

\subsection{experimental conclusion}
In conclusion, we have shown that a high shear is able to generate
in the system a distribution of relaxation time $P(\tau_\alpha)$
independent of the sample history and of shear amplitude. We then
demonstrated that a moderate oscillatory strain both partially
rejuvenate and overage the system. We deduced from that point that
the rejuvenation process is not a simple backwards time
translation of $P(\tau_\alpha)$. It involves a more sophisticated
process during which the shape of the distribution is modified.
However, independently of the imposed perturbation,
$P(\tau_\alpha)$ seems to converge to the same time-dependant
distribution at very long times. In addition, if an oscillatory
strain is applied similarly but at a higher frequency the
overaging process is amplified.

\section{SGR model}
In a previous paper \cite{notrePRL} we emphasized the similarity of this results with the
predictions of the simple trap model \cite{monthus} solved with a step in temperature.
This qualitative agreement reinforces the similarity of shear and temperature on the particles level.
 We also pointed out that a resolution of the Sherington-Kirkpatrick model for spin glasses in the
 transient regime following a temperature step gives the same results \cite{BerthierBouchaud,Berthierprivate}.
  However, the equivalence between shear and temperature increase remains an hypothesis. We present here some
   results on a model where the macroscopic shear is coupled to the microscopic. There are only two of such
    models that we are aware of. One is based on a mode coupling approach and has only be solved in the
     asymptotic regime \cite{Berthier}. The other one is inspired from the trap model and is called the
      Soft Glassy Rheology (SGR) model \cite{PRLLeqHeb,solirevue,Sollitchseul}. We will now solve this
       model for the transient strain history applied to our system.
\par The basis of the model are the following: the system is described by fictive independent
 particles moving in a fixed energy landscape. Only local minima are considered. The particles
  are trapped in wells of depth E from which they escape in a "thermal" like fashion. The escaping
   probability is proportional to $\exp [-E/x]$ where x plays the role of the thermal energy.
    The macroscopic strain is introduced as an external field that shifts the minimum energy levels
     E by an amount $-\frac{1}{2}k l^2(t)$ where $l(t)$ is the local accumulated strain. Thus straining helps
      to hop outside the wells.  In order to
      calculate $l(t)$ the following assumptions are made:

\begin{enumerate}
  \item Each time a particle escape from a trap, it falls in an unconstrained state of depth $E$ with $l=0$.
   The probability of falling in a trap of depth $E$ is proportional to the density $\rho(E)$ of trap $E$.
  \item The strain rate is homogeneous all over the sample.
  \item The two preceding hypothesis allow to define $l(t)$ as the integrated local strain since the last hoping event $t_l$ for the particle: $l(t)=\int_{t_l}^{t}\dot{\gamma}dt$.
  \item The local elastic modulus k is independent of the trap depth.
  \item $\rho(E)$ is exponential: $\rho(E)=\frac{1}{x_g}\exp[-E/x_g]$.
\end{enumerate}
A complete description and justification of the model can be found
in \cite{PRLLeqHeb}.
\newline
We call $P(E,l,t)$, the time-dependant probability for a particle
to be in a well of depth $E$ with a strain $l$. The evolution of
$P(E,l,t)$ in the SGR model reads as:

\begin{widetext}
\begin{equation}\label{eq2}
  \frac{\partial P(E,l,t)}{\partial t}=-\underbrace{P(E,l,t)e^{-(E-\frac{1}{2}kl^2(t))/x}}_{\text{escaping term}}+\underbrace{\Gamma (t)\rho (E)\delta(l)}_{\text{entering term}}-\underbrace{\dot{\gamma}\frac{\partial P(E,l,t)}{\partial l}}_{\text{external advection}}
\end{equation}
\end{widetext}
where the time unit has been set to 1, and with $$\Gamma
(t)=\int_{-\infty}^{\infty}\int_{0}^{\infty
}P(E^{\prime},l',t)e^{-(E-\frac{1}{2}kl'^2(t))/x}dE^{\prime
}dl^{\prime }$$ Each $(E,l)$ state is thus associated with a
relaxation time $\tau_E\propto \exp[(E-1/2kl^2)/x]$. In absence of
shear, the local strain remains always equal to zero, and the
model reduces to the Bouchaud's trap model: for $x>x_g$, a steady distribution for $P(E)$ exists. Oppositely, for $x<x_g$ no steady distribution exist and
$P(E)$ reaches an asymptotic regime with a full
aging behavior. Subsequently $x_g$ is assimilated to a glass
transition temperature. If a constant shear rate is applied, the
system always displays an equilibrium distribution of relaxation
times. The model describes qualitatively well many rheological
features of soft glassy systems
\cite{PRLLeqHeb,solirevue,Sollitchseul}.
\newline
We numerically solve the equation \ref{eq2} by discretizing the equation into 100 energy levels and 100 strain level. We checked that the obtained results do not depend on the numbers of levels we use. We take $k=2$, $x_g=1$ and $x=0.5$. We remark that the values of $\gamma$ can not be compared to that really used in the experiments. We use the initial conditions of a deep quench. The system is supposed to have its equilibrium distribution for $T=\infty$. Hence, $P(E,l,0)=\rho(E)\delta(l)$. The external strain is taken as follows:

\begin{equation}\label{eq3}
  \gamma(t)=\left\{\begin{array}{l}
    0 \text{  if  t}<t_{att} \\
    f(t)\text{  if  }t_{att}\leq t\leq t_{att}+du \\
    0\text{  if  t}>t_{att}+du\
  \end{array}\right.
\end{equation}
f(t) is a function of t that will be specified cases by cases.
\newline We discuss the effect of shear by computing $P(\varepsilon,t)$ with $\varepsilon=E-\frac{1}{2}kl^2$.
Notice that the relaxation time distribution
$P(\log(\tau_E))\propto P(\varepsilon)$. Hence $P(\varepsilon)$
and $ P(\tau_E)$ have the same physical meaning. For sake of
clarity, we will use $P(\epsilon)$ for which all the described
effects are more visible.

\subsection{one step}
Firstly, we exemplify the effect of a single square pulse: $f(t)=\gamma_0$.
\newline We first let the initial distribution $P(E,l,0)=\rho(E)\delta(l)$ evolves between $t=0$ and $t=t_{att}$. We chose $t_{att}$ so that $P(E,l,t_{att})$ has nearly its self-similar shape. In this regime, the $t/t_w$ scaling leads to the following property for the distribution function:
\begin{equation}\label{eq4}
P(E-e,l,t)=P(E,l,t-t_{eff})
\end{equation}
with $t_{eff}=t(1-\exp[-e/x])$ and e is any positive constant.
For $t=t_{att}$, the effect of the step is to shift the states $(E,0)$ to the states $(E,\gamma_0)$.
\newline For $t_{att}<t<t_{att}+du$ the population is now split into two l levels as the particles always jump from
 a $(E,\gamma_0)$
state to a $(E,0)$ one. One can show that, because of equation
\ref{eq4}, the evolution of $P(\varepsilon,t)$ during the step is
the same than that of the unperturbed case but shifted in time by
$t_{eff}=t_{att}(1-exp[-\frac{1}{2}k\gamma_0^2])$.
However,according to equation \ref{eq2} the states $(E,\gamma_0)$
do not age but simply becomes depopulated. Only states $(E,0)$
have a non zero entering rate. Hence there is no possible
self-similar regime for P(E,l,t) until $P(E,\gamma_0)=0$. Hence at
$t=t_{att}+du$ the effect of shear is to shift backwards all the
levels: $(E,\gamma_0)$ states are shifted to $(E,0)$ and (E,0)
states are shifted to $(E,-\gamma_0)$ states. However, the effect
of the strain shift differs from a simple shift in time because
$P(E,l,t)$ is not in its self-similar regime - despite the fact
that $P(\varepsilon)$ has its self similar shape-.
\begin{figure}[ht]
\begin{center}
\includegraphics[width=8cm,keepaspectratio=true]{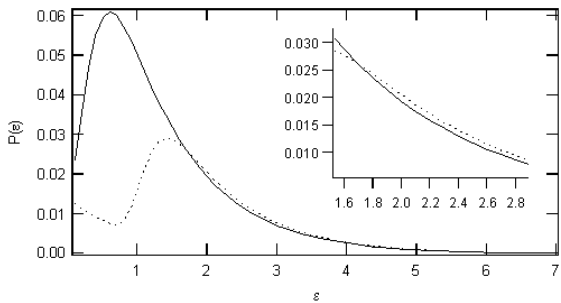}
\caption{Distribution function $P(\varepsilon)$ when no shear is
applied (stray line) and when a strain square pulse $(\gamma_0=)$
is applied. The inset is a zoom on the region where the overaging
is visible. } \label{onestep}
\end{center}
\end{figure}

Fig \ref{onestep}shows that it leads to split of the
$P(\varepsilon)$ into two bumps. Low energies are mainly populated
by the particles that hoped during the square pulse. Conversely,
high energies are mainly populated by particles that did not hop.
But in addition, and as a result of the whole history these levels
become overpopulated by jumps from the low energy levels. Finally,
a splitting of the distribution results from the acceleration of
the kinetics of low energy levels and overaging originates from
this splitting of the energy population.
\newline Discretizing the strain history as very short steps,
one can solve equation \ref{eq2} for any strain history. We will
see in the following section that an oscillatory strain amplifies
deeply the splitting of $P(\varepsilon)$.

\subsection{sinusoidal strain}
\subsubsection{effect of amplitude}
\begin{figure}[ht]
\begin{center}
\includegraphics[width=8cm,keepaspectratio=true]{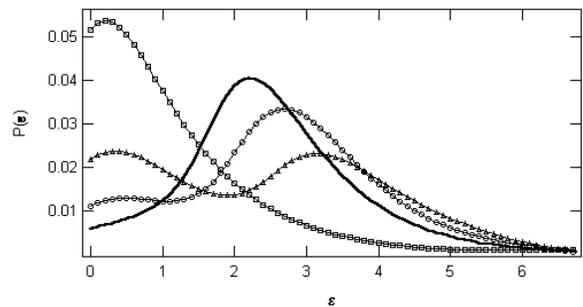}
\caption{Calculated distribution P($\varepsilon$) at 5Hz for
different strain amplitude $\gamma_0=0$ (Bold Line),
$\gamma_0=1.7$ ($\circ$), $\gamma_0=2.5$ ({\tiny $\triangle$}),
$\gamma_0=4.5$ ({\tiny $\square$}). Notice the splitting of the
distribution and the surpopulation of both the short and large
$\varepsilon$ compare to the reference case $\gamma_0=0\%$.}
\label{PdeE}
\end{center}
\end{figure}

In this section we take $f(t)=\gamma_0 sin(\omega t)$ with
$\omega=\frac{2}{5}\pi$. It corresponds to a burst of 10 cycles.
Fig \ref{PdeE} shows $P(\varepsilon)$ 0.1 tu after the burst.

The solid line
corresponds to the unperturbed case: $\gamma_0=0$. The curves lies
in increasing order of strain amplitude $\gamma_0$ from bottom to
top on the right hand-side of the figure. It appears that for
large amplitudes -eg $\gamma_0=4.5$- the system is accelerated.
Indeed, the distribution $P(\varepsilon,4.5)$ lies over the
reference one for small $\varepsilon$ i.e. small relaxation times,
and below the reference one for large $\varepsilon$ i.e. long
relaxation times. It thus corresponds to a case where the system
was rejuvenated. Interestingly, for moderate amplitudes. The
distribution splits into two bumps. Both low and high
$\varepsilon$ lie above the reference curves, whereas the
population of moderate $\varepsilon$ are depleted. In order to
better understand how the splitting of the correlation function
influences the dynamics, we calculated:
$$
C(t_{w}+t,t_{w})=\iint_{0}^{\infty }P(E',l',t_{w})e^{[-t.e^{-(E'-\frac{1}{2}kl'^2)/T}]}dE'dl'
$$
$C(t+t_w,t_w)$ is the probability that a particle has not change
its position between time $t_w$ and $t_w+t$. It is actually
similar to our measured correlation function $g_2(t+t_w,t_w)$.More
precisely, $g_2$ is a monotonic function of $C$, going from 1 to
0, while $C$ is going from 1 to 0. The detailed shape of
${g_2}(C)$ depends on the set-up optical characteristics and is of
no interest in the following. Optically the situation is similar
to the one of structurally evolving foams and is discussed in
\cite{DurianPine}.

\begin{figure}[ht]
\begin{center}
\includegraphics[width=8cm,keepaspectratio=true]{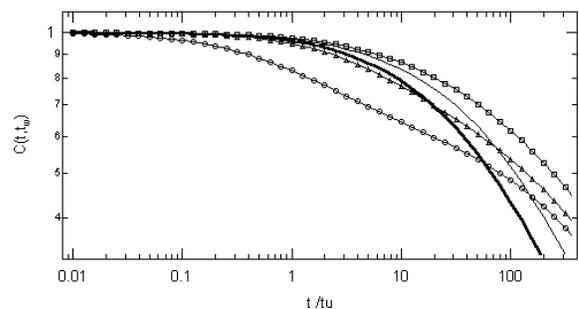}
\caption{Same curve that fig\ref{functcorr} but for the calculated
$C(t_w,t+tw)$ for $t_w$=0.01tu (o),6 tu ({\tiny$\triangle$}), 50
tu ({\tiny$\square$}). The amplitude of the strain is
$\gamma_0=2.5$. } \label{fcorrtheo}
\end{center}
\end{figure}

Fig \ref{fcorrtheo} displays $C(t+t_w,t_w)$ for $\gamma_0=2.5$
and $tw=0.01,6,50$ tu after the burst. The reference curves for
the unperturbed case at $t_w=0.01$ -resp 50 tu- are plotted in
dark -resp grey- solid lines. For $\gamma_0=2.5$, the
correlation function $C(t+t_w,t_w)$ for $t_w=0.01 tu$ (o) starts to
decrease before the unperturbed curve. This is a consequence of
the overpopulation of low energy states. However, the rate of
decorrelation is smaller because the depletion of intermediate
energy state. Since the high energies are overpopulated, the two
curves crosses. At long time, the unperturbed curve lies under the
perturbed one. For longer $t_w$ ({\tiny$\square$}), the low energy states
have aged and the shape of the correlation function is dominated
by the population of the high energy. The correlation function for
$\gamma_0=0.06$ looks 'older' than the reference one. Notice the
excellent qualitative agreement between fig \ref{fcorrtheo} and
fig \ref{functcorr}.

\subsubsection{effect of frequency}
In this section we now focus on the effect of the burst frequency
upon the dynamics. We keep the same initial conditions but have
$\omega$ varying from $2\pi /50$ to $2\pi$. It corresponds to a
number of oscillations ranging from 1 to 50. As explained in the
previous section we calculated $C(t+t_w,t_w)$ for different $t_w$
after the burst.

\begin{figure}[ht]
\begin{center}
\includegraphics[width=8cm,keepaspectratio=true]{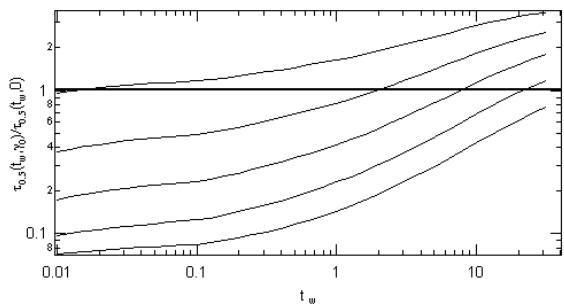}
\caption{Calculated $\tau_{0.5}(t_w,\gamma_0)/\tau_{0.5}(t_w,0)$
for different frequencies of oscillation. The amplitude $\gamma_0$
is 10. The frequencies are 0.1Hz, 0.2Hz, 0.3Hz, 0.4Hz, 0.5Hz from
bottom to top. Notice that the overaging effect is more sensitive
to the frequency for the model than for the experiment. }
\label{effreqtheo}
\end{center}
\end{figure} 

Following the analysis procedure we used in the experimental
section, we define $\tau_{0.5}(t_w,\gamma_0)$ such that
$C(\tau_{0.5}+t_w,t_w)=0.5$. Figure\ref{effreqtheo} displays
$\tau_{0.5}(t_w,\gamma_0)$ as a function of $t_w$. One can see
that it qualitatively predicts the same frequency behavior than
that displayed by our sample -see fig\ref{freqexp}. Actually
oscillatory shear strain, by mixing in strain and thus in
$\varepsilon$ the population, considerably accelerates the
dynamics of the small energy levels, while it does not affect the
high energy levels. Thus increasing the frequency $f(t)$  - for
the same duration of application - mainly accelerates the low
energy dynamics, by making the strain mixing more efficient. Thus
the oscillatory strain is qualitatively similar to the square
pulse strain. But the amplitude of the splitting in the
$\varepsilon$ distribution and thus of the overaging considerably
increases with  the frequency.

\subsection{discussion}
 We want first to emphasize the following point: overaging is a notion associated with a
\textbf{transient} perturbation and a comparison with a reference
case. It requires not only an analysis of the average relaxation
time but of the global shape of $P(\varepsilon,t)$. Overaging comes from the overpopulation of the long relaxation times after the shear application. However, the experiments and the model show that the short relaxation times are simultaneously overpopulated. This is what we called rejuvenation. Because of the simultaneous occurrence of theses two phenomena on different time scale, an exact analysis can only be performed if the complete distribution $P(\varepsilon)$ is studied. However, the experimental time window is limited and the system can thus appear rejuvenated or overaged depending on the relaxation times that we can probe. We now wonder if a regime where rejuvenation alone can be achieved for any time scale after a perturbation $\gamma_0$ and a duration du. The experiment duration are too limited to give a satisfactory answer. However we can make the following remarks about the SGR model. In this model, aging occurs because for $x<x_g$ the escaping rate $\Gamma_{out}(E)$ is inferior to the entering rate $\Gamma_{in}(E)$ for high enough energy wells. $\Gamma_{out}(E)$ is proportional to $\exp[-E/x]$ and thus depends only on the on the well's depth. Oppositely, $\Gamma_{in}(E)$ is proportional to $\Gamma(t)$ and thus depends on the whole distribution $P(E)$. Let us first examine the population of the level (E,0) before the shear such that $$ E \gg E_c=\frac{1}{2}k\gamma_0^2+x\ln(du)$$. When a shear is applied, $\Gamma_{out}(E)$ is simply multiplied by a factor $\exp[\frac{1}{2}kl^2/x]$ during a time du. Hence escaping probability of this population will remain nearly zero despite the shear application.
 Theses particles will remain in the states $(E,0)$ after the shear has been applied. However, for these energies, the entering rate is increased because of the overpopulation of  small energies. Overaging comes thus from the fact that the escaping rate remains unchanged whereas the entering rate is increased for large energies. Hence overaging in the SGR model occurs for any finite $\gamma_0$ and any finite duration. In this case a complete rejuvenation is impossible.
However, the assumption that the trap stiffness is independent of the well depth is only a first step assumption. One could imagine that the deeper the well the stiffer its elastic modulus. One could for instance assume that:
$$k=k_0+\kappa E$$ where $\kappa$ is a constant of proportionality.
Making this assumption first changes the instantaneous elastic modulus $<k>$ of the model. It is no longer $k_0$ but now
$$<k>=k_0+\kappa <E>$$
For $x>x_g$ a stationary distribution
of P(E) exists and we find $<k>= k_0+\kappa \frac{x_g*x}{x-x_g}$.
In the limit of infinite temperature $<k>$ tends towards $k_0+\kappa x_g$. However, when $x<x_g$, the average
energy $<E>$ is proportional to $\log(t_w)$ in the asymptotic
regime. The instantaneous elastic modulus has thus a logarithmic
dependance with the age of the system as observed in many systems
\cite{CaroFranc,cloitre1,Bonnlapo,Cipelonion}.
An other change affects the escaping rate.
$\Gamma_{out}$ is now proportional to $\exp[-[\frac{(1-\frac{1}{2}\kappa l^2)}{x}E-\frac{1}{2x}k_0 l^2]]$.
The strain now plays a role similar to a temperature change. In this case, the value of $E_c$ now reads:
$$E_c=\frac{\frac{1}{2}k\gamma_0^2+x\ln(du)}{1-\frac{1}{2}\kappa \gamma_0^2}$$
One can thus define a critical strain
$\gamma_{0c}=\frac{2}{\kappa}$ for which $E_c$ is infinite. Hence a
complete rejuvenation ca be achieved in a finite amount of
time. From an experimental point of view, the distinction between
the two approximations by a rejuvenation experiment are not yet
concluding. Some other experiments are being performed and the
consequences of this modification for the stiffness will be
carefully discussed elsewhere.

\section{conclusion}
In conclusion, we have shown that an oscillatory strain can act on
a glassy colloidal suspension in a dual fashion. It can rejuvenate
it by erasing all its past memory. It can also overage it by
accelerating the aging process. Both processes comes from an
acceleration of the rearrangement rates during a transient time.
The dominating effect depends on the amplitude and duration of the
shear. We showed that the effect is qualitatively well explained
by the SGR model. This model was solved for a realistic strain
history. We showed that the distribution of relaxation time was
separated into two bumps, one corresponding to rejuvenation and
the other one to overaging. The influence of this two bumps on the
position autocorrelation function was studied. We showed that the
model and the experiments are highly comparable. We believe that
this phenomena is not a special feature of colloidal systems but
can be observed in polymer - as indicated by preliminary results
performed in our lab - or in spin glasses.

\end{document}